\documentclass{epl}
\usepackage{amssymb}

\title{Breakdown of step-flow growth in unstable homoepitaxy}
\shorttitle{Breakdown of step-flow growth...}
\author{J. Kallunki \inst{1} \and J. Krug \inst{2,3}}
\institute{
\inst{1} Laboratory of Physics, Helsinki University of
Technology, P.O. Box 1100, FIN-02015 HUT, Espoo, 
Finland  \\
\inst{2} Fachbereich Physik, Universit\"at Duisburg-Essen, 
45117 Essen, Germany \\ 
\inst{3} Institut f\"ur Theoretische Physik, Universit\"at zu 
K\"oln, Z\"ulpicher Strasse 77, 50937 K\"oln, 
Germany\footnote{Present and
permanent address.}
}


\begin{document}

\maketitle

\begin{abstract} 
Two mechanisms for the breakdown of step flow growth, in the sense of the 
appearance of steps of opposite sign to the original vicinality, are studied
by kinetic Monte Carlo simulations and scaling arguments. The first mechanism
is the nucleation of islands on the terraces, which leads to mound formation
if interlayer transport is sufficiently inhibited. The second mechanism is the
formation of vacancy islands due to the self-crossing of strongly meandering
steps. The competing roles of the growth of the meander amplitude and the
synchronization of the meander phase are emphasized. 
The distance between vacancy islands along the step direction appears
to be proportional to the \emph{square} of the meander wavelength.

\end{abstract}

\section{Introduction}

Growing surfaces are commonly divided
into two classes, vicinal or singular 
\cite{Pimpinelli98,Michely03,Politi00}. 
On vicinal surfaces 
the density of atomic steps, originating from the miscut of the
sample, is high, so that most of the atoms arriving on the
surface attach to these steps and the nucleation
of adatom islands on the terraces may be neglected. The
crystal then grows in the step-flow mode, 
simply by propagation of the steps.
As no new steps appear, all the steps have the same sign, and
the surface profile remains monotonic in the direction of vicinality.
Even if the step morphology is unstable against step meandering, 
the growth may be considered as a two-dimensional problem.  

It has been observed however in experiments 
\cite{Tejedor99,Maroutian_phd}
as well as in computer simulations \cite{Tejedor99,Krug95,Rost96} 
that the step-flow mode is only metastable. 
Eventually steps having a sign opposite to the original
vicinality appear on the surface. This destroys the two dimensional
character of the growth and leads to the formation of genuinely
three-dimensional structures.

Steps of opposite sign may originate from two different
mechanisms: either from the nucleation of adatom islands on the
terraces, or through the formation of vacancy islands. Island
nucleation on the terraces is severely suppressed by the 
steps as they collect the deposited adatoms, keeping the 
adatom concentration low. Due to fluctuations, however, some nucleation
events do occur. Nucleation of  islands on the terraces alone
is not enough to destabilize the step-flow growth, as the
islands collide with the propagating steps and merge with them.
The step-flow growth is destabilized only if new islands are
nucleated, on the average, on top of the terrace islands 
before the latter are incorporated into the steps.

Vacancy islands can be formed on the terraces due to the
meandering of the steps. During growth the atomic steps often
do not remain straight, but form a meander pattern \cite{Maroutian_phd,Maroutian99}
either due to the Bales-Zangwill (BZ) instability \cite{Bales90}, 
or due to a Kink-Ehrlich-Schwoebel effect (KESE)
\cite{Pierre-Louis99,Ramana99}. As the meander amplitude
becomes large in the course of time, it is plausible that a 
step might cross itself forming thus a closed loop, \emph{i.e.}
a vacancy island. Before turning to the detailed analysis of this
effect, which forms the main subject of this Letter, we briefly
discuss destabilization due to the nucleation of islands.   
   
\section{Mound formation through terrace nucleation}

A common criterion for step-flow growth is that the
step distance $\ell$ should be smaller than the nucleation 
length $\ell_D$, the distance between the nucleation 
centers on a singular surface; for irreversible aggregation
$\ell_D \sim (D/F)^{1/6}$, where $D$ is the adatom diffusion
coefficient and $F$ is the deposition flux \cite{Pimpinelli98,Michely03}.
When $\ell \ll \ell_{D}$ essentially all atoms go to the
existing steps and nucleation events are very rare.  
Due to fluctuations in the deposition beam, however, 
nucleation of adatom islands occurs also during step-flow growth.
When $\ell$ is not too much smaller than $\ell_D$, 
islands are continuously 
nucleated on the terraces and incorporated into the
steps. The nucleation length is 
altered by the presence of the steps and becomes
$
\tilde{\ell}_D \approx (D/F) \ell^{-5}$ 
parallel to the steps \cite{Bales96,Rusanen03}. 
Coalescence of the terrace islands with
the steps also affects the step morphology 
\cite{Rusanen03}.

Nucleation of islands on the terraces is not enough to
launch mound formation. It is also required that, \emph{on average},
a new island will be nucleated on top of the terrace island
before it is incorporated into the step,
because this process must be repeated several times when a
mound starts to form. 
Nucleation on top of islands
is enhanced by an Ehrlich-Schwoebel (ES) barrier,
which prevents the descent of adatoms across steps \cite{ES66}.
The strength of this effect depends on the ratio of the ES-length
$\ell_{ES} = D/D'$, where $D'$ is the interlayer hopping rate, to 
the island radius $R$ \cite{Michely03,Politi00}. 
When $\ell_{ES} \ll R$ second layer nucleation 
on a singular surface does
not occur until the first layer islands have started to coalesce, 
at $R \approx \ell_D$, and
the initial growth proceeds layer by layer; on a vicinal surface 
this implies that second layer
nucleation is preempted by incorporation of the 
first layer islands.

Destabilization of step flow growth by mound formation thus
requires a strong ES effect, in the sense that $R \ll \ell_{ES}$ at
the time of incorporation. To obtain a quantitative criterion, 
we start from the expression 
$
\omega(R) = \pi^2 F^2 R^5/2D'
$
for the nucleation rate on top of an island of radius $R$, which
holds for strong ES barriers and irreversible nucleation \cite{Krug00}.
The probability that nucleation has occurred on the island
within a time $t$ after its creation is then 
\begin{equation}
\label{nuc_number}
p_{\mathrm{nuc}}(t) = 1 - \exp \left[- \int_{0}^{t} \omega(R(s)) {\mathrm d}s
\right], 
\end{equation}
where $R(t)$ is the island radius at time $t$.
Mound formation requires that $p_{\mathrm{nuc}}$ grows to order unity,
say $p_{\mathrm{nuc}} = 1/2$, during the time $t^\ast$ that the step
needs to travel a distance $\ell/2$, and hence to engulf the island.
As the step velocity is $v_{\mathrm{step}} = F \ell$, we have
$t^\ast = 1/(2 F)$. We further assume  that half of the
material landing on a terrace segment of length $R$ contributes
to the growth of the island, so that
$R(t) = F \ell t/2$. Evaluating (\ref{nuc_number}) and
setting $p_{\mathrm{nuc}}(t^\ast) = 1/2$, we find that mound formation
occurs when  
\begin{equation}
\label{lC}
\ell > \ell_c \approx 4.4 \times (D'/F)^{1/5}.
\end{equation}
The length scale $\ell_c$ has the same dependence on the growth parameters
as the radius of the top terrace of a mound growing on a singular surface
\cite{Michely03,Krug00}, though the dimensionless prefactor is somewhat
larger. Note that since for strong ES barriers $\ell_c \ll \ell_D$, 
the condition (\ref{lC}) does not contradict the conventional step flow
criterion $\ell \ll \ell_D$. 

This analysis assumes that steps are straight and equidistant, 
and that the nucleation events on the terraces are uncorrelated. 
In reality the step train may suffer from growth instabilities
either in the step direction (\emph{step meandering})
\cite{Politi00,Maroutian99,Bales90,Pierre-Louis99,Ramana99,Kallunki02} or in the direction of the
vicinality (\emph{step bunching}) \cite{Pimpinelli98};
sometimes the two instabilities even coexist \cite{Neel03}.
In the case of 
step bunching the large terraces forming between the bunches 
are prominent sites for island nucleation and consequently for the formation
of mounds. Also step meandering leads to a non-uniform terrace width. 
In an in-phase meander the terrace width becomes locally smaller than 
the initial terrace width $\ell$ \cite{Kallunki02}, but at sites where the
phase-shift between consecutive steps is large, large terraces
are formed. This is discussed in more detail in the following section.
In growth experiments on Cu(1 1 12) a fairly regular
array of mounds was observed \cite{Maroutian_phd}.  This 
suggests that the mounds do not form independently of each other.

\section{Vacancies and craters} 

\begin{figure}
\begin{center}
\includegraphics[height=6.8cm, angle = 0]{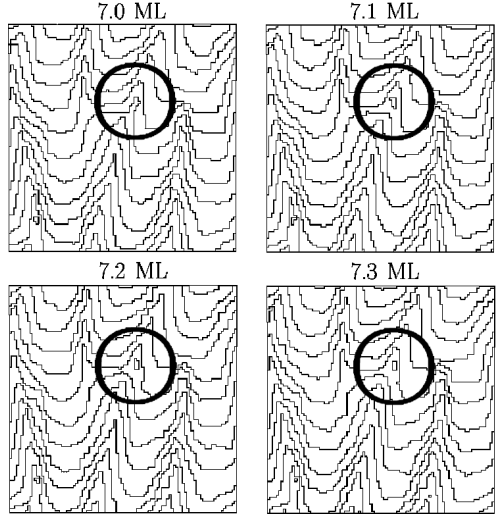}
\caption{Consecutive step configurations  
of the SOS model at time intervals of 0.1 ML after deposition 
of 7 ML.  The flux is
$F=1.0 \, \mathrm{ML/s}$ and the 
system size $100 \times 100$ .
}
\label{vac_form}
\end{center}
\end{figure}

The formation of vacancy islands was studied using
kinetic Monte-Carlo simulations of a 
standard SOS model on a square lattice.    
The elementary processes are the deposition 
of atoms at rate $F$ and the hopping of adatoms 
to nearest neighbor sites with a rate 
$
\nu=\nu_0 \exp(-E_a/k_B T)$.
The activation energy
$E_a$ includes 
an energy barrier $E_{S}$ for diffusion on a flat terrace, 
a contribution $E_{nn}$ for each lateral in-plane bond,
an additional energy barrier
$E_{BB}$ for breaking a bond, and an ES-barrier
$E_{ES}$, which is implemented through the change
in the number of next-nearest neighbors in the planes
above and below the hopping atom. 
The bond breaking barrier $E_{BB}$ serves to enhance step edge diffusion
compared to detachment from the step.  
We used rectangular lattices with $L_x$ ($L_y$) sites along (perpendicular to)
the steps and periodic boundary conditions in the $x$-direction.
The nominal step spacing was $\ell = 10$.
For a detailed description 
of the model we refer to \cite{Kallunki02}.

The activation barriers were set to the values 
$E_{S} = 0.35$~eV, $E_{nn} = 0.15$~eV,
$E_{BB} =0.2$~eV and
$E_{ES} = 0.2$~eV. The temperature  was $T=400$ K and the 
diffusion pre-factor \mbox{$\nu_0=4.17 \times 10^{12} \, \mathrm{s}^{-1}$.}
For this choice of parameters the steps undergo a meandering
instability due to the Kink-Ehrlich-Schwoebel
effect (KESE) \cite{Pierre-Louis99,Kallunki02}, leading to 
wavy steps with a wavelength
\begin{equation}
\label{lambda_M}
\lambda_M \approx (12 D_s/F\ell)^{1/4},
\end{equation}
where $D_s$ is the diffusion rate along a straight step;
within our model, the activation energy for edge diffusion is
$E_S + E_{nn}$. 
In the KESE instability
the steps start to meander with random phase shifts, 
in contrast to the Bales-Zangwill instability, where the
steps meander collectively and the step modulations are in-phase from the
outset \cite{Kallunki02}. Due to the ES barrier the phase correlation
increases during time\footnote{The step meander approaches the in-phase mode even
without an ES-barrier, but the ordering process is
much slower and the meander wavelength seems to increase
during ordering \cite{remark}.}, 
as the steps have a tendency to follow the
leading step \cite{Kallunki_phd}. Even if the phase correlation 
increases locally, however, globally the steps remain incoherent, leading to
occasional
large phase shifts between subsequent steps where two domains
of (relatively) coherent meander are joined, see Fig \ref{vac_form}.
At these sites the strong meander easily leads to formation
of a closed step loop, as a step crosses itself. This 
process is shown in Fig. \ref{vac_form}. For the parameters used in 
the simulations the critical length (\ref{lC}) 
$\ell_c \approx 320$, 
so the formation of mounds due to island
nucleation can safely be disregarded. 
The average distance of islands in the step direction
is $\tilde{\ell}_D \approx 100$ even for the largest value
used for the flux $F$. In the simulations
adatom islands were never observed. 

\begin{figure}
\begin{center}
\twoimages[height=4.5cm, angle=0]{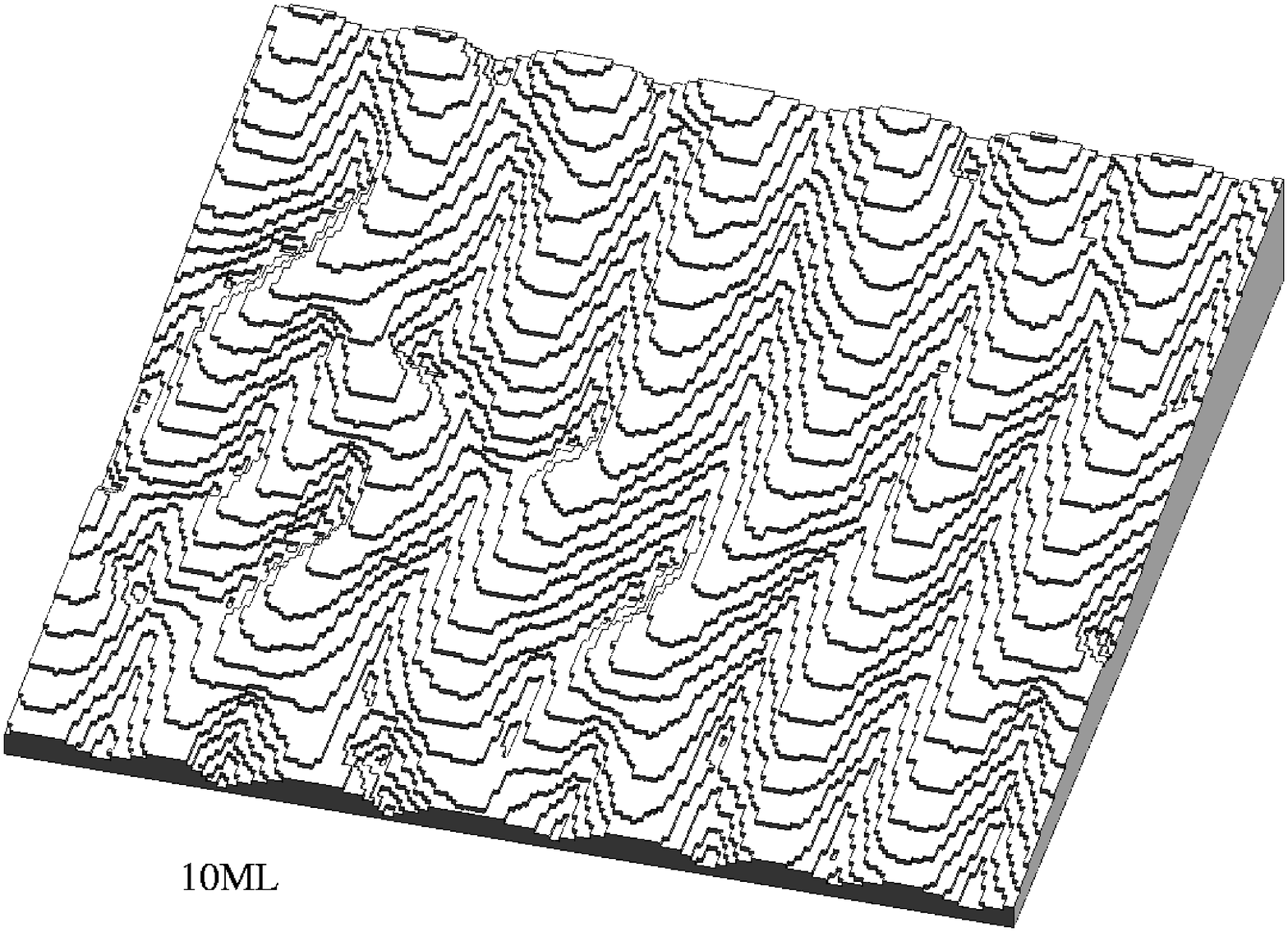}{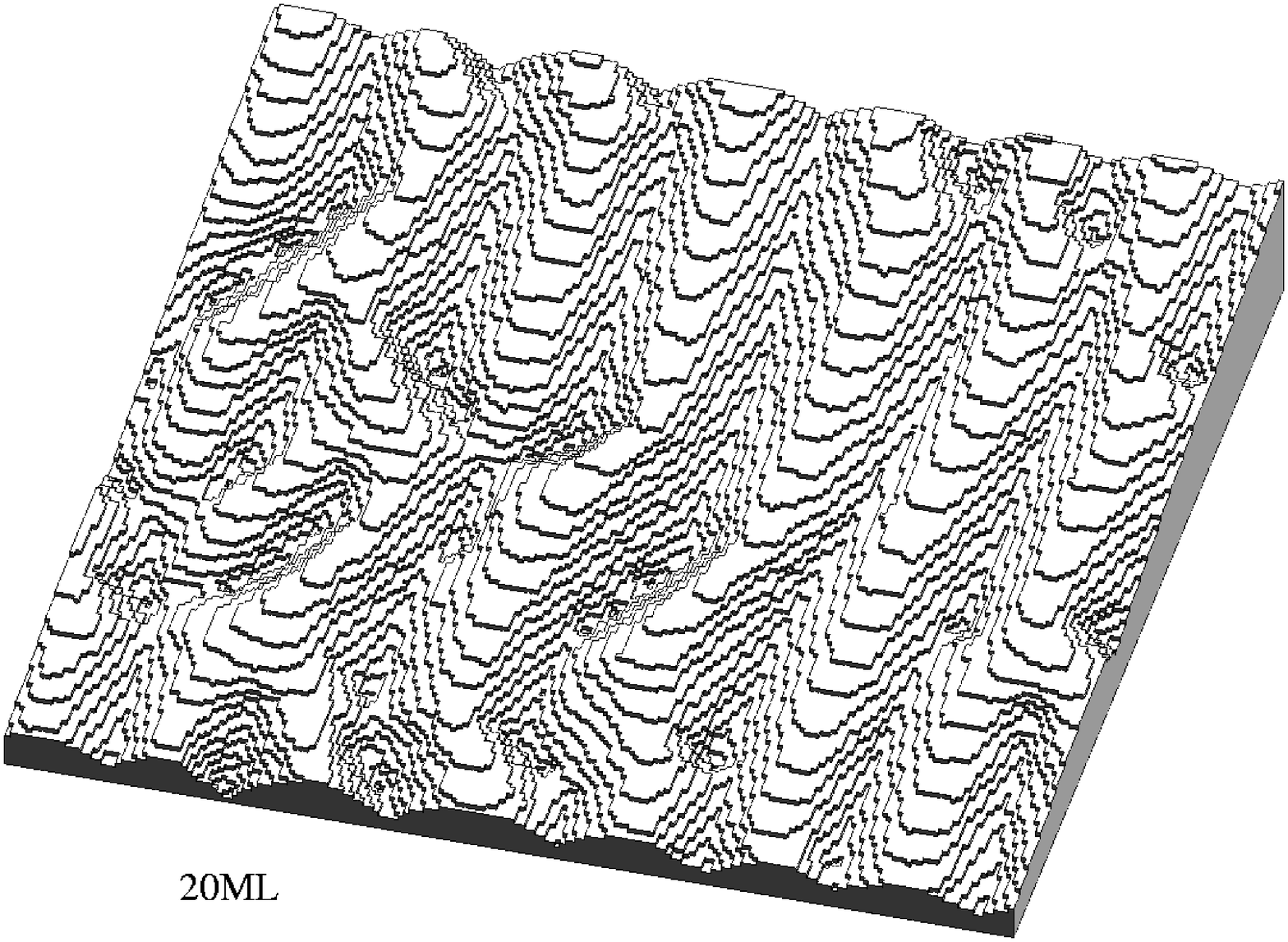}
\twoimages[height=4.5cm, angle=0]{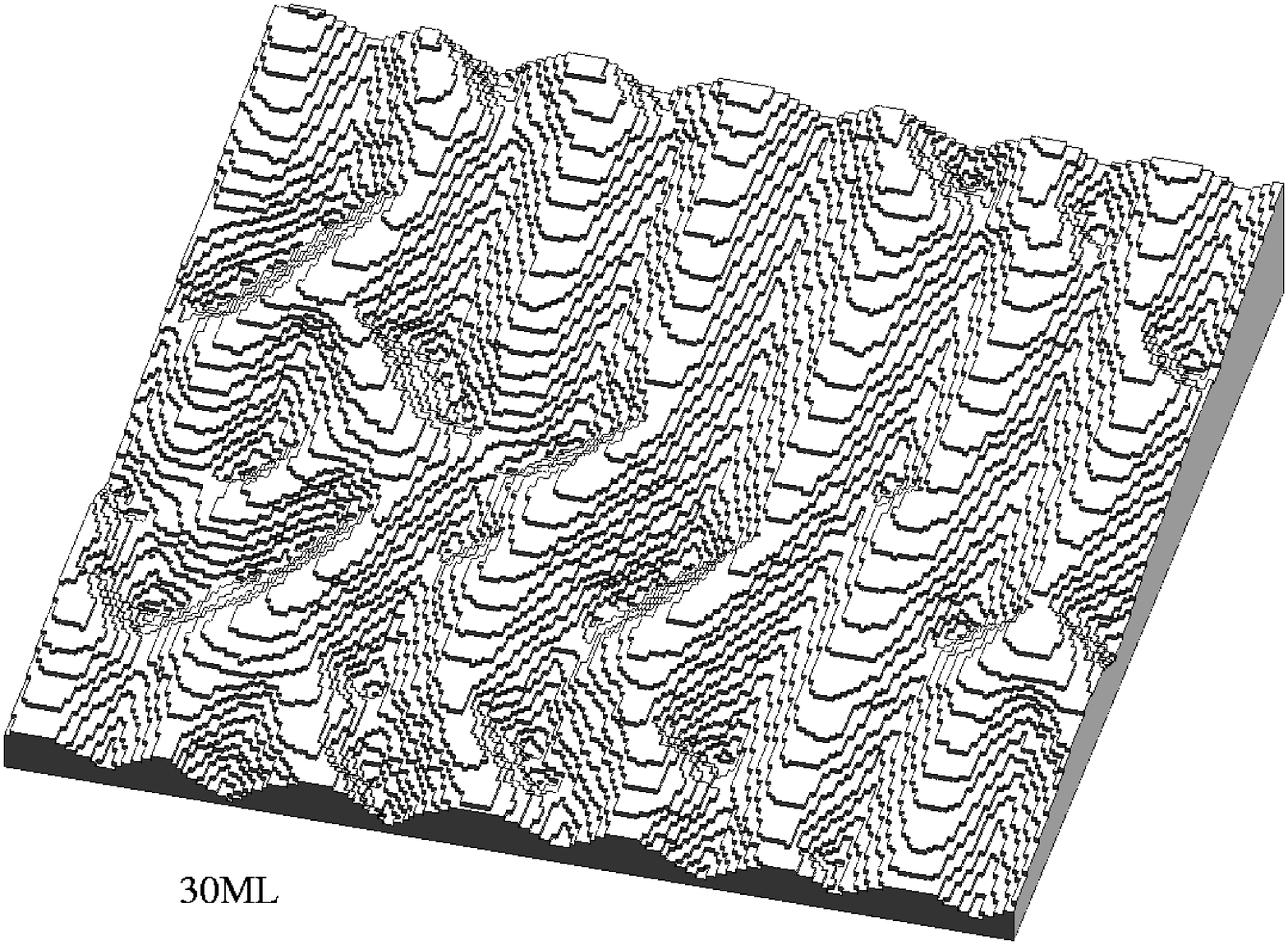}{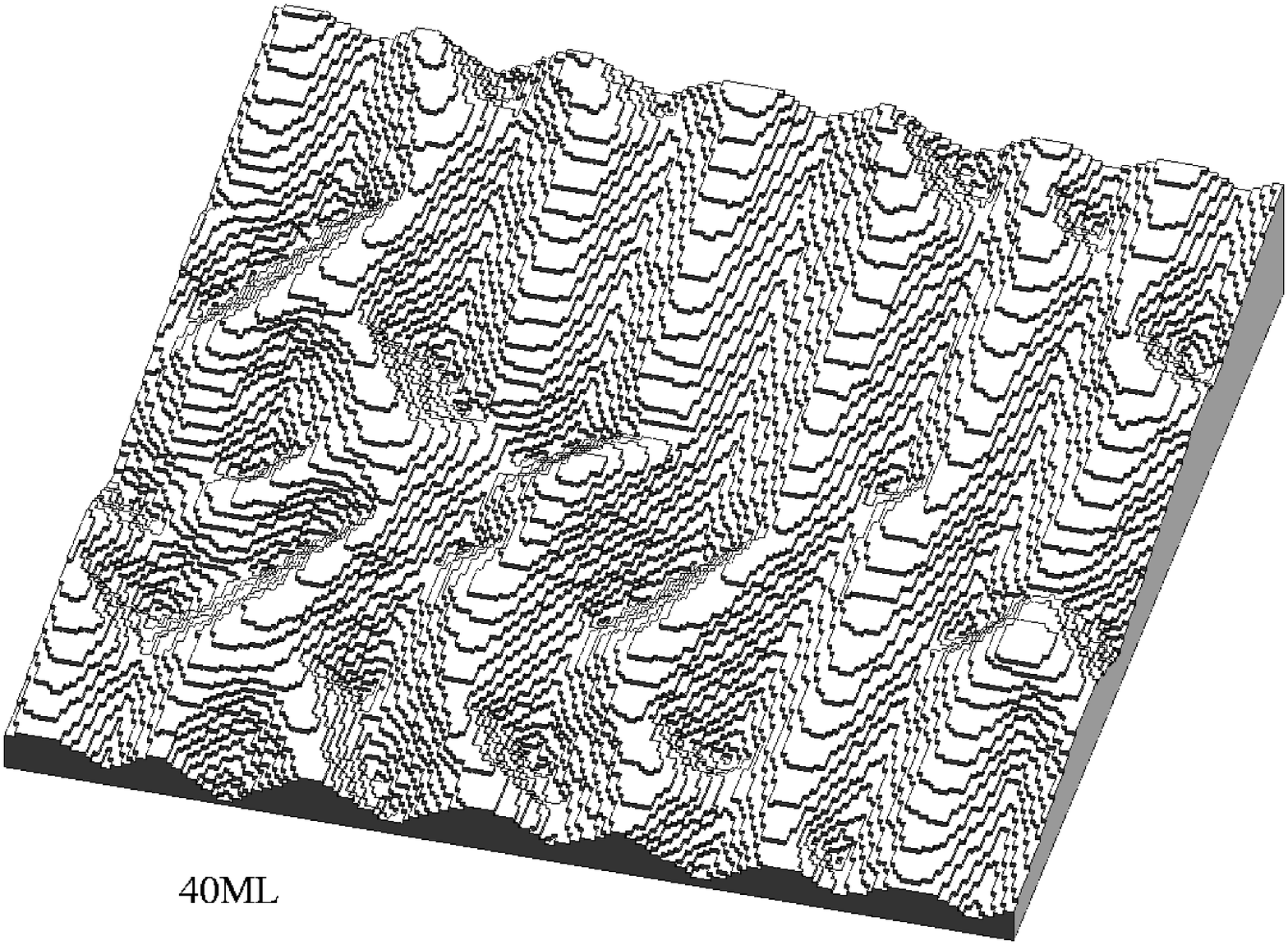}
\caption{Evolution of the surface morphology after the appearance
of vacancy islands. Step pinning leads to the formation of
deep craters. 
After deposition of 40 ML islands form on 
terraces which are severely deformed due to the neighboring
craters. The flux is $F=2.0 \mathrm{ML/s}$ and  the system size
$200 \times 200$ .}
\label{crater_grow}
\end{center}
\end{figure}

Unless the vacancy island created through the self-crossing of a step
is rapidly filled, it acts as a pinning center for the following steps.
Consequently the pinned steps then also form closed loops around
the pinning center, and a deepening crater appears. The 
development is seen in the snap shots in Fig. \ref{crater_grow}.
It should be noted that the filling of vacancies is severely
hampered by the ES-barrier. During further growth,
the uphill surface current induced by the ES-barrier \cite{Michely03,Politi00} leads
to the deepening, lateral growth and coalescence of craters.
As seen in Fig. \ref{crater_grow}, the ripple pattern is now
severely distorted and large terraces between 
neighboring carters appear. Eventually islands are nucleated on these
large terraces, launching the formation of a mound.

Creation of mounds has been observed also earlier in simulations
of a similar SOS model \cite{Tejedor99,Rost96}. In the
previous works the meandering of the steps was due to
the BZ-instability, rather than the KESE mechanism
which is relevant for this work. Also in the earlier works
the mounds were seen form preferentially at the
sites where the phase coherence between the steps
is poor \cite{Tejedor99,Rost96}. 

\section{Statistics of vacancy formation}
\begin{figure}
\begin{center}
\twofigures[width=6cm]{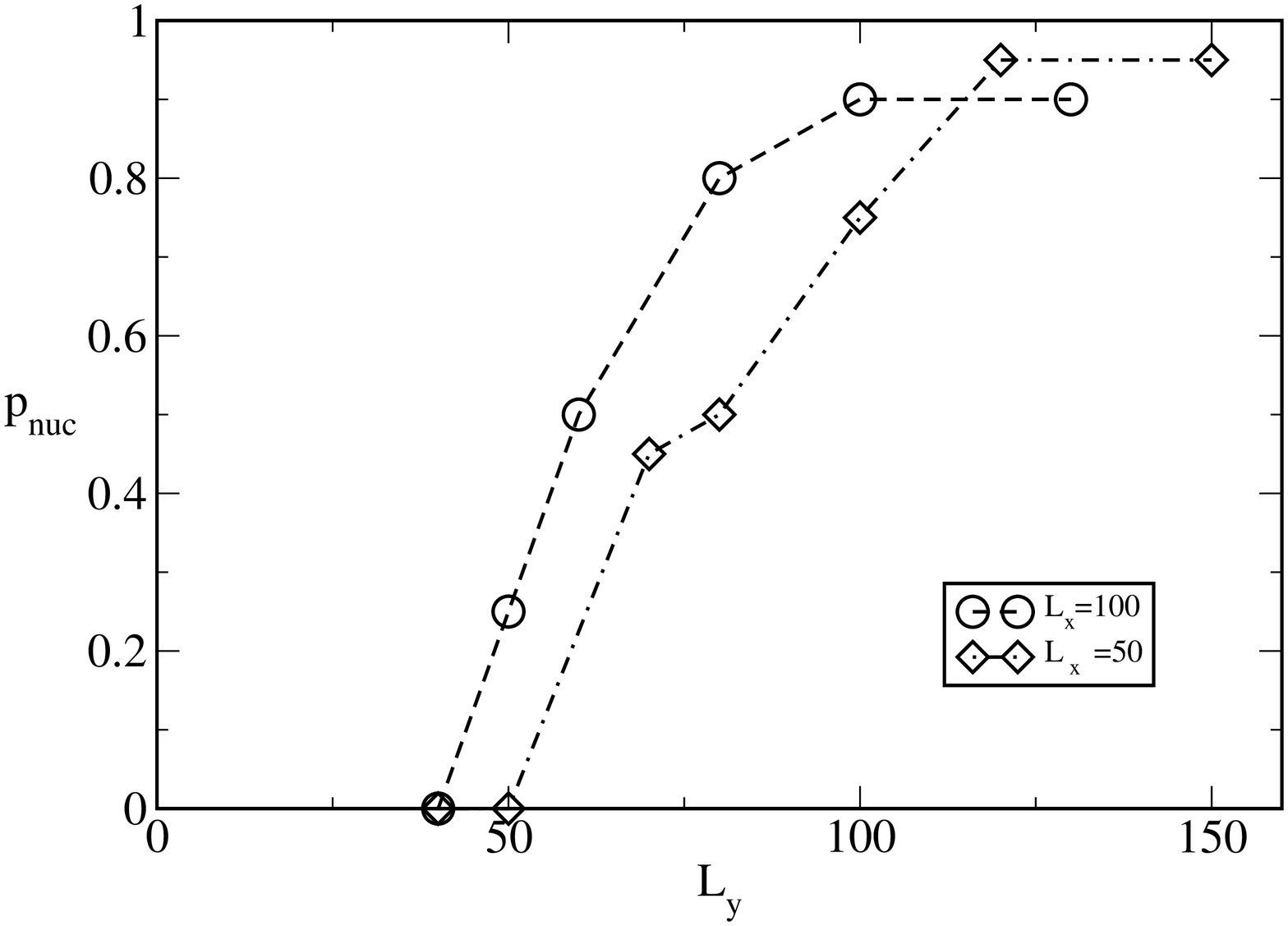}{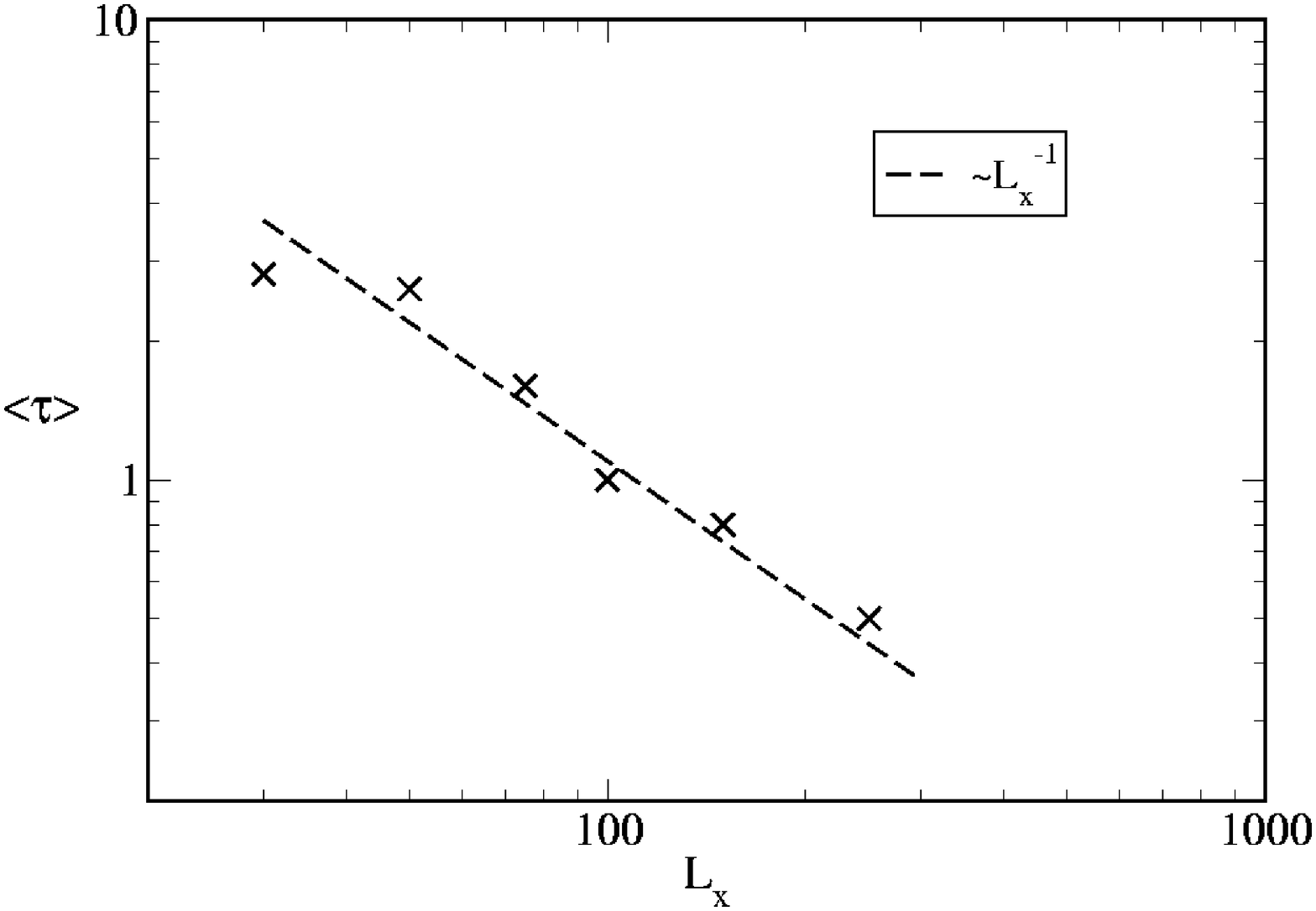} 
\caption{Probability for the formation of a vacancy island in a small
system. Results are obtained from 20 independent runs with a deposition
flux $F=5.0$ML/s.}
\label{pnuc-Ly}
\caption{Mean waiting time $\langle \tau \rangle$  before a 
vacancy is created for $L_y = 100$ and varying $L_x$,
obtained from 20 independent runs for each point with  $F=5.0$ML/s. 
Dashed line shows the expected scaling
$\langle \tau \rangle \sim L_{x}^{-1}$.}
\label{tx-Lx}
\end{center}
\end{figure}

We have seen above that the
formation of a vacancy island requires a large amplitude meander and
a large phase shift between consecutive steps. However,
as the meander amplitude grows in time, the phase correlation
also improves. The spatial distribution of the vacancies 
results from the competition between
these two effects. To acquire good 
statistics for the vacancy separation, simulations
of extremely large system sizes would be necessary.
Here we focus instead on the statistics of the appearance
of the \emph{first} vacancy island in the system.

Vacancy islands were identified as 
closed step loops in the SOS configuration.
This method does not 
differentiate between vacancy and adatom islands, but in the
parameter regime of our simulations
no island nucleation takes place on the terraces. 
The loops were searched by first finding the contour
lines separating the lattice in parts where the surface height
$h(i,j)$ is higher or lower than some reference height $h_{ref}$. 
By varying the reference height $h_{ref}$ between the highest
and the lowest value of the height in the configuration,
all steps are found. After finding a step it was checked whether
the same point appears in the contour twice, indicating a closed
loop. Steps having length 4 were neglected as they mark adatoms.

\begin{figure}
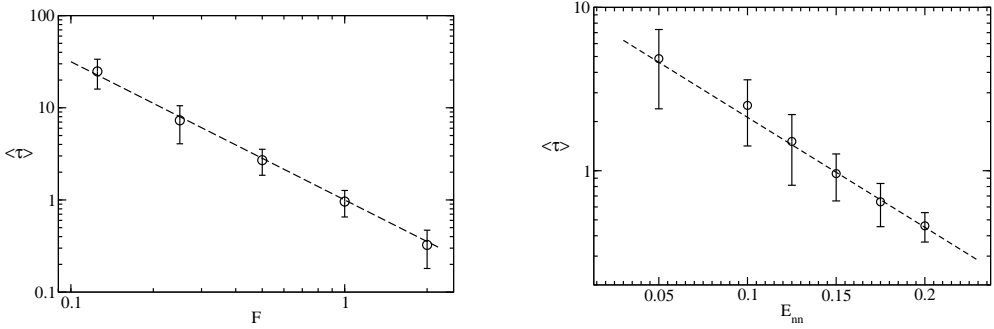

\begin{center}
\twoimages[width=6cm]{time_first_vs_F.eps}{time_first_vs_Enn.eps}
\caption{Mean waiting time $\langle \tau \rangle$ for the creation
of a vacancy island as a function of the
deposition flux $F$ and the bond energy $E_{nn}$. 
The results are obtained from 20 independent runs for each point.
The dashed lines are $\langle \tau \rangle \sim F^{-3/2}$ (left panel) 
and  $\langle \tau \rangle \sim \exp(-E_{nn}/2 k_B T)$ (right panel). 
A slightly different set of parameters was used; the Schwoebel barrier
was $E_{ES} = 0.15$~eV, temperature $T = 375$~K and the deposition
flux $F=1.0$ML/s (right panel). Also the diffusion pre-factor had a 
different value $\nu_0=1 \times 10^{12} 
\, \mathrm{s}^{-1}$, so that the ratio $D/F$ remained almost the same. }
\label{tau}
\end{center}
\end{figure}

To study the
vacancy separation in the direction of the vicinality
it was investigated whether vacancies form at all in systems containing  
a relatively small number of 
steps, $L_{y}/\ell = 4 - 15$. For small
$L_y$ the phase coherence propagates through 
the whole system before the meander amplitude becomes large enough
for loop formation and
the final configuration is a perfect in-phase step train.
For larger $L_{y}$ vacancies may form
before the meander phases have synchronized.  
Thus varying the system size $L_{y}$, keeping the step
distance fixed, an estimate for the distance
between the vacancies in the vicinal direction is
obtained. Results from such simulations 
are shown in Fig. \ref{pnuc-Ly}. The probability for the creation of
a vacancy changes quite rapidly from zero to one as the number of
steps in the system increases. The data
also show a weak dependence on the step length 
$L_{x}$, which we however expect to saturate at large $L_x$. 
Based on Fig. \ref{pnuc-Ly}, we estimate
that the distance between vacancies in the $y$-direction is of the order
of 50 under the conditions of the simulations.

The most natural candidate for the vacancy distance in the
step direction is the meander wavelength $\lambda_{M}$. 
To study this length scale,
the average waiting time
$\langle \tau \rangle$ before the appearence of the first vacany, 
starting with a train of perfectly straight steps, was measured.
The waiting time is inversely proportional to the number $N_{vac}$ of
potential vacancy formation sites. For fixed $L_y$ we expect that  
$N_{vac} \sim L_{x}/\lambda_{M}$, and hence  
$\langle \tau \rangle \sim \lambda_M/L_{x}$.
Figure \ref{tx-Lx} confirms that $\langle \tau \rangle \sim L_{x}^{-1}$,
showing that there is a constant vacancy formation probability per
unit step length.
To test the relation $\langle \tau \rangle \sim \lambda_M$, we have conducted
simulations with various values of the growth flux 
$F$ and the bond energy 
$E_{nn}$, which control the meander wavelength (\ref{lambda_M}) 
\cite{Kallunki02}. Results for $L_x=200$, $L_y = 100$ are plotted in Fig. \ref{tau}.
Surprisingly, the data are consistent with  
the relationship $\langle F \tau
\rangle \sim \lambda_M^{-2}$, which indicates that the number of
potential vacancy sites scales as $\lambda_M^{-2}$ rather than as
$\lambda_M^{-1}$ We have no explanation to offer at present. 
Qualitatively, we note that 
the deposition flux and the bond energy also affect
the growth rate of the meander amplitude and the ordering time scale 
into the in-phase mode, and hence changing these parameters alters
also the vacancy separation in the vicinal direction in addition
to the meander wavelength.

\section{Conclusions and outlook}
\label{Conclusions}

In this paper we have studied two mechanisms for the destabilization
of step flow growth in the presence of an Ehrlich-Schwoebel effect.
We have derived a quantitative criterion for destabilization
due to mound formation, and presented extensive simulation results
for destabilization due to vacancy island formation mediated by
strong step meandering. In the latter case we have found
evidence for the appearance of a new length scale, proportional to 
the square of the meander wavelength, which determines the 
distance between vacancy islands along the step direction.

An important insight is that the vacancy formation process is controlled
by the competition between the growth of the meander amplitude and the
increase of phase correlations between consecutive steps. 
This suggests that the behavior may be rather different when step
meandering is driven by the Bales-Zangwill-instability, rather than by 
the KESE mechanism considered here, 
because in the former case phase correlations
are present from the very outset. It may also be of
interest to extend the present study to systems without an explicit
ES-barrier. On the one hand, in such a system the phase correlations
develop much more slowly \cite{remark}, favoring the formation
of vacancy islands; on the other hand, in the 
absence of the ES-barrier the deepening and coarsening
of craters would also be slowed down.

\acknowledgements

We are grateful to H.J. Ernst, M. Kotrla, T. Maroutian and 
O. Pierre-Louis for discussions and correspondence. 
This work was supported in part by DFG within SFB 237 \emph{Unordnung und grosse
Fluktuationen}, and by Volkswagenstiftung.



\end{document}